%% file: bare_jrnl.tex
\documentclass[journal]{IEEEtran}
\ifCLASSINFOpdf
\else
\fi
\usepackage{mathtools}
\usepackage{color}
\usepackage{algorithm}
\usepackage[noend]{algpseudocode}

\usepackage{times}
\usepackage{epsfig}
\usepackage{graphicx}
\usepackage{amssymb}
\usepackage{romannum}
\usepackage{color}

\usepackage{amsmath}
\usepackage{breqn}
\usepackage{multirow}
\usepackage{subfigure}
\usepackage{xurl}

\usepackage{textcomp}

\begin{document}
\title{Calibrated Bagging Deep Learning for Image Semantic Segmentation: A Case Study on COVID-19 Chest X-ray Image}

 \author{Lucy Nwosu,~Xiangfang Li,~Lijun Qian,~Seungchan Kim,~Xishuang Dong
        % <-this % stops a space
\thanks{L. Nwosu, X. Li, L. Qian, S. Kim and X. Dong are with the Department of Electrical and Computer Engineering and the Center for Computational Systems Biology, Prairie View A\&M University, Texas A\&M University System, Prairie View, TX 77446, USA. Email: lnwosu@pvamu.edu, xili@pvamu.edu, liqian@pvamu.edu, sekim@pvamu.edu, xidong@pvamu.edu}% <-this % stops a space
}

\maketitle

\begin{abstract}
Severe acute respiratory syndrome coronavirus 2 (SARS-CoV-2) causes coronavirus disease 2019 (COVID-19). Imaging tests such as chest X-ray (CXR) and computed tomography (CT) can provide useful information to clinical staff for facilitating a diagnosis of COVID-19 in a more efficient and comprehensive manner. As a breakthrough of artificial intelligence (AI), deep learning has been applied to perform COVID-19 infection region segmentation and disease classification by analyzing CXR and CT data. However, prediction uncertainty of deep learning models for these tasks, which is very important to safety-critical applications like medical image processing, has not been comprehensively investigated. In this work, we propose a novel ensemble deep learning model through integrating bagging deep learning and model calibration to not only enhance segmentation performance, but also reduce prediction uncertainty. The proposed method has been validated on a large dataset that is associated with CXR image segmentation. Experimental results demonstrate that the proposed method can improve the segmentation performance, as well as decrease prediction uncertainties.

\end{abstract}


% \fi
\IEEEpeerreviewmaketitle

\section{Introduction }
\label{sec1}
\input{Introduction}

\section{Methodology}
\label{sec2}
\input{Method}

\section{Experiment }
\label{sec4}
\input{Experiment}

\section{Related Work }
\label{sec5}
\input{Relatedwork}

\section{Conclusion and Future Work}
\label{sec7}
\input{Conclusion}

\section*{Acknowledgment}
\label{acknowledgement}
This research work is supported in part by the Texas A\&M Chancellor's Research Initiative (CRI), the U.S. National Science Foundation (NSF) award 1736196, and by the U.S. Office of the Under Secretary of Defense for Research and  Engineering (OUSD(R\&E)) under agreement number FA8750-15-2-0119. The U.S. Government is authorized to reproduce and distribute reprints for Governmental purposes notwithstanding any copyright notation thereon. The views and conclusions contained herein are those of the authors and should not be interpreted as necessarily representing the official policies or endorsements, either expressed or implied, of the U.S. National Science Foundation (NSF) or the U.S. Office of the Under Secretary of Defense for Research and Engineering (OUSD(R\&E)) or the U.S. Government. 

\ifCLASSOPTIONcaptionsoff
  \newpage
\fi


\bibliographystyle{IEEEtran}
\bibliography{Reference}

\end{document}

%% file: Introduction.tex
%\section{Introduction}

Severe acute respiratory syndrome coronavirus 2 (SARS-CoV-2) causes coronavirus disease 2019 (COVID-19) which was first identified in 2019 in Wuhan, Central China~\cite{wang2020comprehensive}. It is spreading globally, resulting in more than 458 million confirmed infections and 6 million deaths, and causing huge economic loss. Although global economics seems to be recovered gradually, early and accurate tests of this disease such as reverse transcription-polymerase chain reaction (RT-PCR), antigen tests, and medical imaging tests must be improved to be ready for future pademics~\cite{ozturk2021classification, zhang2021diagnosis}. Compared to RT-PCR tests, medical imaging tests such as chest X-ray (CXR) and computed tomography (CT) are more effective and efficient~\cite{pan2020time, wong2020frequency}, especially for severe patients, which is of great help to physicians. For instance, in Italy, the United States, and China, the majority of serious COVID-19 cases have been identified through the manifestation characteristics in CT images~\cite{kumar2020spreading}. Therefore, effective extraction of COVID-related information on  medical images will play an important role to fight against a new round of pandemic caused by COVID mutated variant~\cite{liang2020handbook}.

Deep learning (DL) played an important role in promoting COVID-related information extraction by COVID-19 infection region segmentation and disease classification through analyzing CXR and CT data~\cite{ghoshal2020estimating, gozes2020rapid}. Compared with CT images, CXR images are easier to obtain in radiological inspections. Currently,  most of DL models, especially convolutional neural networks (CNN), were employed to classify entire CXR images to detect COVID-19 cases~\cite{wang2020covid, wang2020abnormal}. For example, Hemdan~\textit{et al.} proposed COVIDX-Net to assist radiologists to diagnose COVID-19 based on CXR features~\cite{hemdan2020covidx}. It integrated various deep convolutional neural networks (DCNNs) models with different structures, such as DenseNet201~\cite{huang2017densely},  Xception~\cite{chollet2017xception}, and MobileNetV2~\cite{sandler2018mobilenetv2}. Sethy~\textit{et al.}  integrated different DCNNs models with a support vector machine (SVM) classifier to recognize COVID-19~\cite{sethy2020detection}.  In addition, to address the shortcomings of training data, Castiglioni~\textit{et al.} employed transfer deep learning techniques for COVID-19 classification, where the pretrained models were built based on ResNet on ImageNet datasets~\cite{castiglioni2020artificial}. Ioannis~\textit{et al.} comprehensively evaluated transfer learning based COVID-19 classification by investigating 5 DCNN models, including VGG19, MobileNetV2, Inception, Xception, and InceptionResNetV2~\cite{apostolopoulos2020covid}.  Similarly, Narin~\textit{et al.} applied 3 typical pretrained DCNN models (i.e., ResNet50, InceptionV3, and InceptionResNetV2) to classify COVID-19 on a small-scale CXR dataset~\cite{narin2021automatic}. Moreover, Lucy~\textit{et al.}~\cite{nwosu2021semi} developed two-path semi-supervised deep learning model to implement COVID-19 classification by using huge amounts of unlabeled data.

Compared with CXR classification, CXR semantic segmentation is a more challenging task that is to classify each pixel into predefined classes~\cite{paul2020generalizability} to recognize region of interests (ROIs) on CXR images, where a few previous work explored this task~\cite{teixeira2021impact, chakraborty2021covid, bhattacharyya2022deep}.  However, prediction uncertainty of DL models for this task has not been comprehensively investigated since most of DL models focus on performance improvement on this task such as increasing detection accuracy. For safety-critical applications like medical image processing, the prediction uncertainty of DL models is a key evaluation metic on reliability of model predictions since high prediction uncertainty means low prediction reliability.  For example, for COVID-19 applications, applying uncertain predictions to clinical processes would result in disastrous consequences such as missing servere COVID cases or delayed treatments.

\begin{figure*} [ht]
 	\centering
	\includegraphics[width=.9\linewidth]{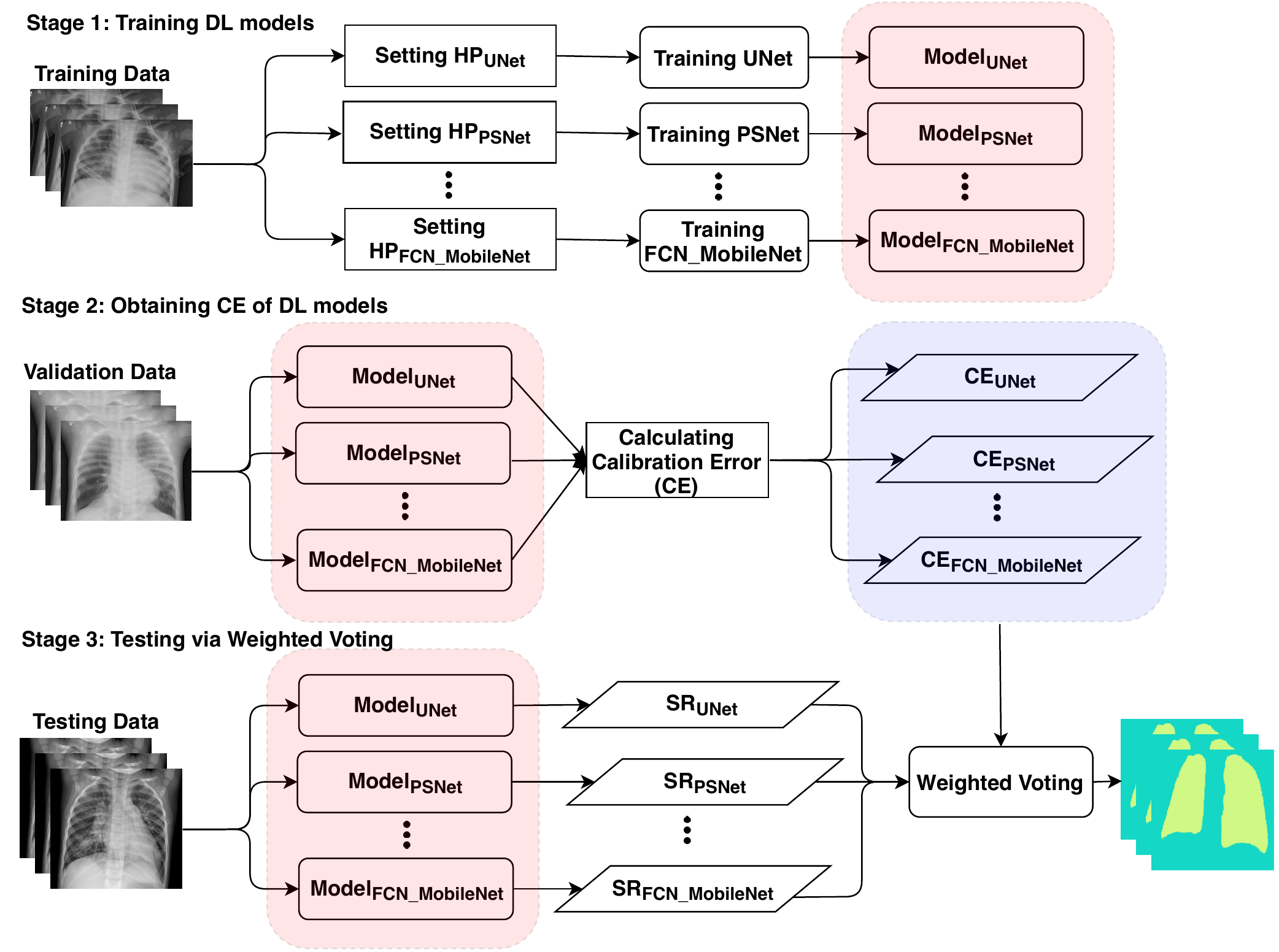}
	\caption{Flow of building and testing calibrated bagging deep learning based on calibration error (CE). $SR$ denotes segmentation result generated by individual deep learning model.  }
	\label{Fig2_baggingcalibrate}
\end{figure*}

This paper proposed a novel ensemble deep learning model that integrates bagging deep learning~\cite{ganaie2021ensemble} and model calibration~\cite{guo2017calibration} to enhance performance of semantic segmentation, as well as reduce prediction uncertainty. It includes three stages: 1) training multiple state-of-the-art DL models such as fully convolutional networks (FCN)~\cite{howard2017mobilenets}, FCN combined with ResNet~\cite{he2016deep}, FCN combined with MobileNet~\cite{howard2017mobilenets}, PSPNet~\cite{zhao2017pyramid}, and UNet~\cite{ronneberger2015u} on training CXR datasets; 2) Calculating calibration errors to measure prediction uncertainties of these DL models on validation CXR datasets, where expected calibration error (ECE) and maximum calibration error (MCE)~\cite{guo2017calibration} are employed to measure the prediction uncertainties; 3) Implementing calibrated bagging deep learning with weighted voting, where the weight of each DL model is inversely proportional to the calibration error. The proposed model is validated on a large-scale CXR dataset to examine its effectiveness. Experimental results demonstrate that the proposed method not only enhances the performance of semantic segmentation, but also improves the prediction certainty on CXR data.

The contributions in this study are below.

\begin{itemize}
\item We systematically compared performance of various state-of-the-art DL models on semantic segmentation on COVID-19 CXR data with different evaluation metrics. Moreover, the prediction uncertainty of these DL models were investigated by measuring expected calibration error (ECE) and maximum calibration error (MCE).

\item We implemented a novel ensemble deep learning model based on model calibration and bagging deep learning, which is to calibrate bagging deep learning models through weighted summation of predictions generated by individual models.  The proposed approach is easily implemented and scalable to various tasks.

\item We validate the proposed method with semantic segmentation on a large COVID-19 CXR dataset based on different evaluation metrics. Experimental results demonstrates its effectiveness on improving performance and prediction certainty for semantic segmentation. 

\end{itemize}

%% file: Method.tex
% Methodology

The proposed method is built based on calibration error~\cite{murphy1967verification, crowson2016assessing, jiang2012calibrating} and bagging deep learning~\cite{ganaie2021ensemble} to enhance image segmentation with higher prediction certainty.

\subsection{Calibration Error}

%The prediction reliability of machine learning models is critical for high risk applications such as medical diagnosis~\cite{crowson2016assessing, jiang2012calibrating} and self-driving~\cite{sayin2021crowd}, which can be formulated as model calibration~\cite{murphy1967verification} that refers to the process of adjusting model parameters to make prediction confidence to be accurate estimation of the probability of the correct prediction~\cite{guo2017calibration}. 

The expected calibration error (ECE) and the maximum calibration error (MCE) are proposed to measure the quality of uncertainty for machine learning models in terms of prediction accuracy~\cite{naeini2015obtaining}, which is critical for high risk applications such as medical diagnosis~\cite{crowson2016assessing, jiang2012calibrating} and self-driving~\cite{sayin2021crowd}.
%Suppose a machine learning model is perfectly calibrated given a class $y$ with the true probability $p$.

%\begin{equation}
%P(\hat{y} = y|\hat{p} = p) = p
%\end{equation}

%where $p \in [0,1]$  and class labels $y \in \{0, . . . , k\}$. It means that when the prediction probability $\hat{p}$ is equal to true probability $p$, the prediction $\hat{y}$ is the same as the ground true $y$. Furthermore, the difference between the prediction confidence $P$ and the true probability $p$ is defined as calibration error, that is to estimate model uncertainty. 

\begin{itemize}
\item Expected Calibration Error (ECE). It estimates the calibration error in expectation values with three steps: 1) Discretizing the prediction probability region into a fixed number of bins; 2) Assigning each predicted probability to one of these bins; 3) Calculating the difference between the fraction of predictions in the bin that are correct (accuracy) and the mean of the probabilities in the bin (confidence) by
\begin{equation}
ECE = \sum_{k=1}^{K}\frac{n_k}{N}\*|acc(k) – conf(k)|
\end{equation}
where $n_{k}$ is the number of predictions in bin $k$, $N$ is the total number of samples predicted, and $acc(k)$ and $conf(k)$ denonte the accuracy and confidence in the bin $k$, respectively. It is a weighted average of differences of accuracy vs confidence in these bins.

\item Maximum Calibration Error (MCE). It measures an upper bound of ECE that is the maximum difference between accuracy and confidence over all predictions across all bins. 

\begin{equation}
MCE = \max_{k=1}^{K}\*|acc(k) – conf(k)|
\end{equation}
\end{itemize}

In summary, MCE measures the largest calibration gap across all bins, whereas ECE measures a weighted average of all gaps. Both MCE and ECE equal 0 if the model is perfectly calibrated.

\subsection{Bagging Learning}

\begin{figure} [ht]
 	\centering
	\includegraphics[width=.9\linewidth]{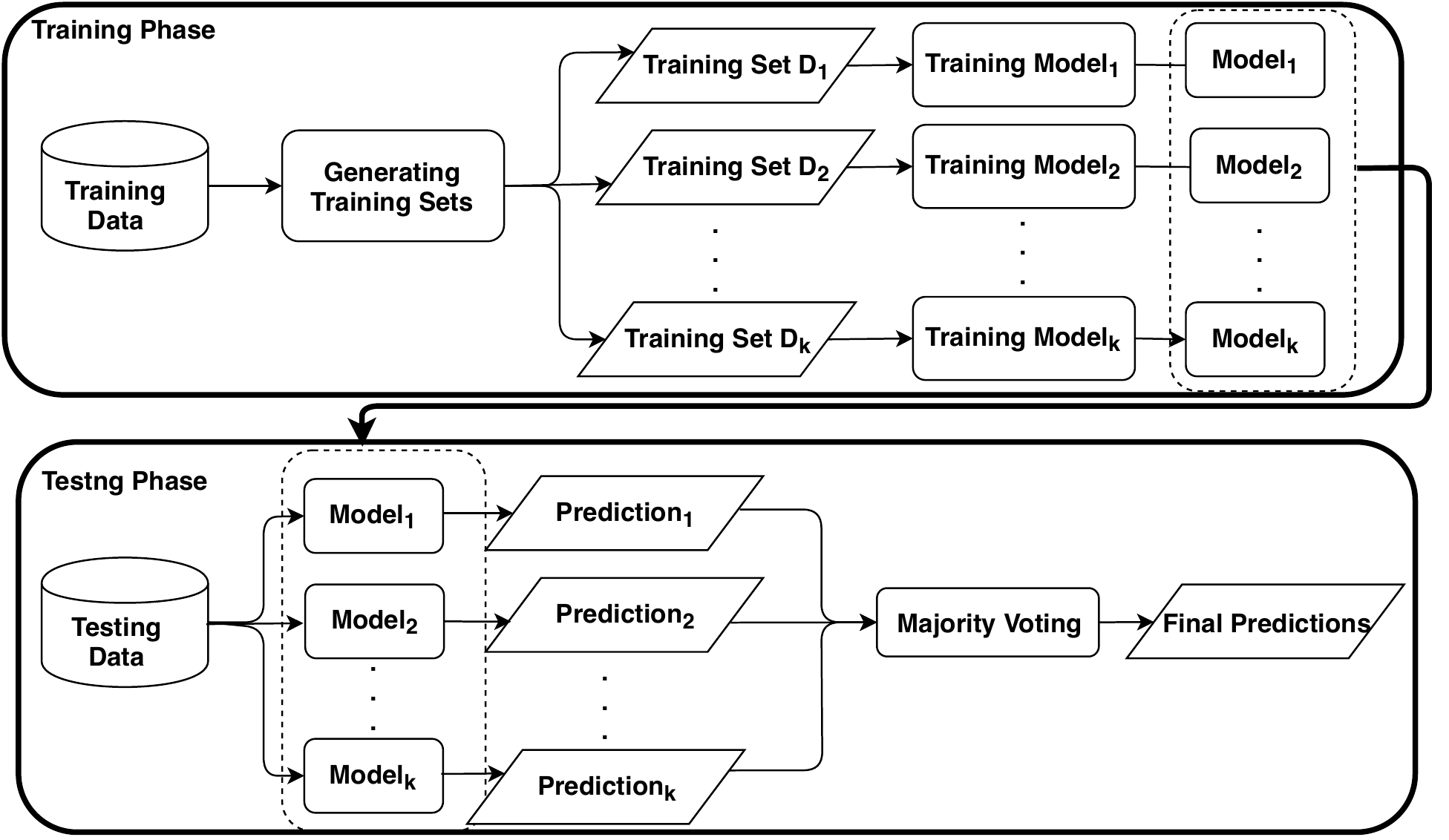}
	\caption{Diagram for building a bagging deep learning model. The model can be different deep learning models such as convolutional neural networks (CNN) and recurrent neural networks RNN) for different applications.  }
	\label{Fig1_bagging}
\end{figure}

Ensemble deep learning combines several individual deep models to improve generalization performance through various ensemble strategies such as bagging and boosting, which integrates the advantages of both deep learning  and ensemble learning~\cite{ganaie2021ensemble}. Bagging (or bootstrap aggregating) generates a series of independent subsets from training data to build multiple individual predictors to build an ensemble model~\cite{breiman1996bagging}. In detail, it generates the bagging samples and passes each bag of samples to base models to build multiple predictors. Then, it is to combine predictions of these multiple predictors with specific strategies such as majority voting. Fig.~\ref{Fig1_bagging} presents a diagram for building and testing bagging deep learning with majority voting, where multiple training sets can be generated by sampling with or without replacement.

\subsection{Proposed Model}

We proposed a calibrated bagging deep learning model to enhance generalization performance as well as reduce prediction uncertainty for COVID-19 semantic segmentation that is to recognize lung region of CXR images. Fig.~\ref{Fig2_baggingcalibrate} presents the flow for building the proposed approach. It includes three stages: 1) training various state-of-the-art deep learning models such as UNet~\cite{ronneberger2015u}, PSPNet~\cite{zhao2017pyramid}, and MobileNet~\cite{howard2017mobilenets}, on an identical training data for COVID-19 image segmentation models, which differs from the standard strategy for bagging learning that is to generate a bag of training sets on original training data; 2) Estimating calibration error (CE) for these different models. First, it is to complete COVID-19 semantic segmentation on validation data by running these DL models to obtain prediction probabilities and accuracy. Then, it calculates CE including ECE and MCE to evaluate uncertainties of these DL models; 3) Testing via weighted voting bagging deep learning. We perform calibrated bagging prediction on testing data through implementing weighted voting, where the weights are built with CE of these DL models. It assumes that lower CE of DL models means higher certainty of these DL models. Moreover,  DL models with the higher certainty are assigned with more weights. Therefore, we define the weight of $i$th model  as  $\frac{1}{CE_{i}}$, where $CE_{i}$ is the calibration error for $i$th model. For COVID-19 semantic segmentation, it is to classify each pixel into either Lung or NonLung. If $\sum^{Lung} \frac{1}{CE_{i}} > \sum^{NonLung}\frac{1}{CE_{j}}$ for one pixel in a CXR image, this pixel is classified as Lung, otherwise,  Non-Lung.

More details on building calibrated bagging deep learning is illustrated in Algorithm~\ref{alg:bagging}, where $M$ denotes the number of the-state-of-art deep learning models involved.

\begin{algorithm}
\caption{Building calibrated bagging deep learning}
\label{alg:bagging}
\begin{algorithmic}[1]
\Require 
{Training set $D_{training}$ and validation set $D_{val}$ }
\Ensure {Calibrated bagging deep learning}
\For{$m \gets 1$ to $M$}                    
        \State {Setting hyper-parameter (HP) for $DL_{m}$}
        \State {Training $DL_{m}$ on  $D_{training}$}
        \State {Calculating $CE_{m}$ of $DL_{m}$ on $D_{val}$ }
\EndFor
\State \Return {DL models $DL = \{DL_{1}, DL_{2}, ..., DL_{M}\}$ and corresponding $CE = \{CE_{1}, CE_{2}, ..., CE_{M}\}$ }
\end{algorithmic}
\end{algorithm}

%% file: Experiment.tex
\newcommand{\photo}[1]
{
    \includegraphics[width=2.5cm]{#1}
}

\subsection{Dataset}

\begin{figure}[h!]
 \begin{center}
\begin{tabular}{cc}

	\photo{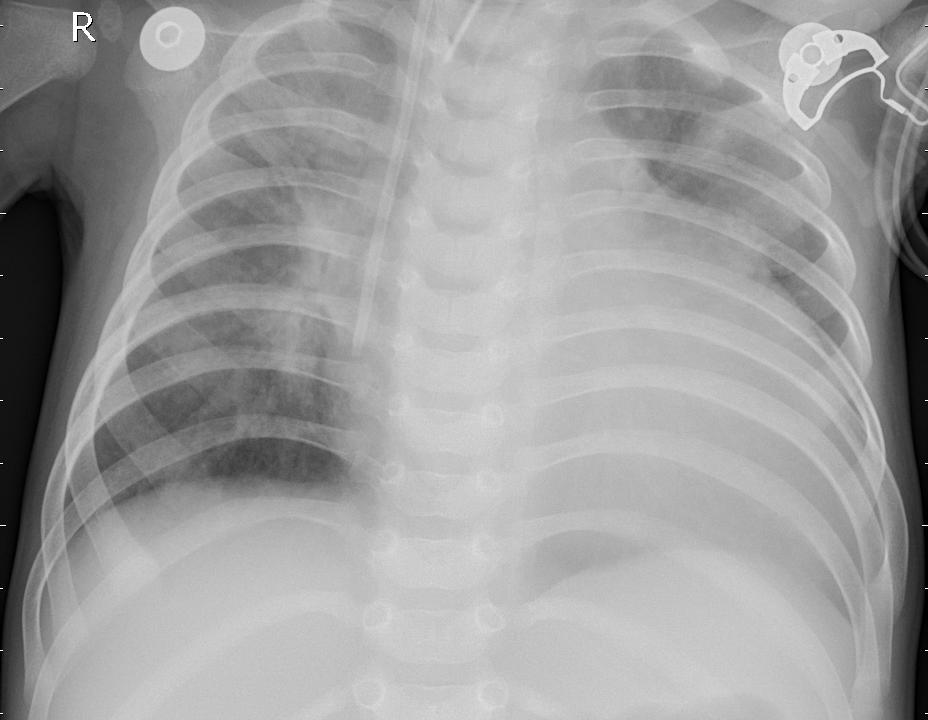} & \photo{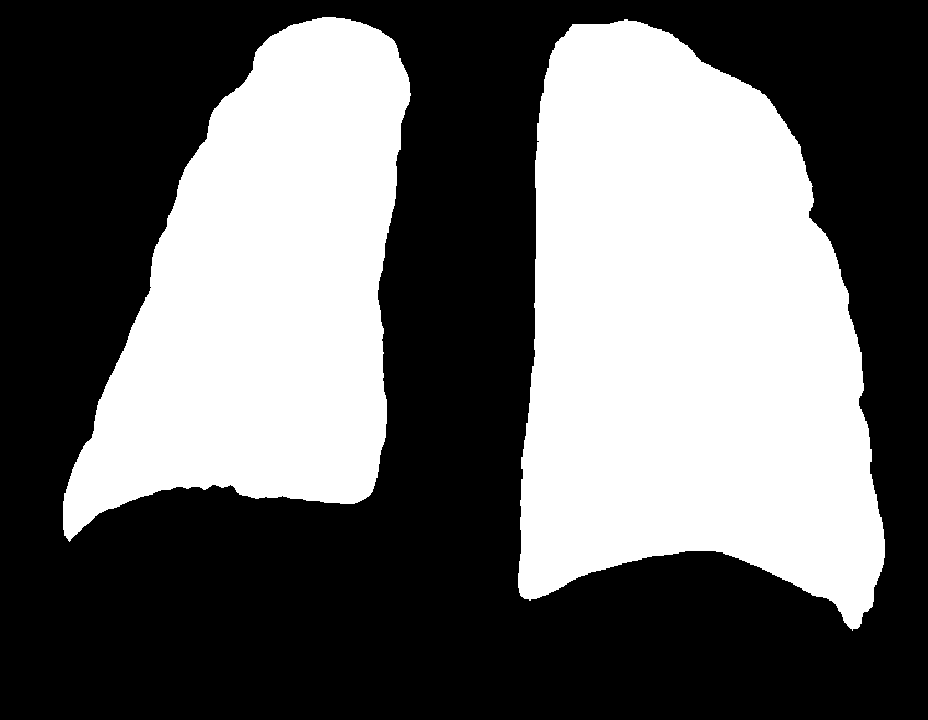}  \\
	(a) original image &  (b) ground truth  \\

\end{tabular}
 \end{center}
 \caption{An example of CXR image and corresponding ground truth.}
 \label{Fig_data}
\end{figure}

We employed COVID-19 chest X-ray dataset\footnote{\url{https://github.com/v7labs/covid-19-xray-dataset}} to validate the effectiveness of the proposed method. It includes $6,402$ images of AP/PA chest x-rays/CT scan with pixel-level polygonal lung segmentations. Each image has a corresponding ground truth with two ``Lung" segmentation masks (rendered as polygons, including the posterior region behind the heart), where the masks include most of the heart, revealing lung opacities behind the heart which may be relevant for assessing the severity of viral infection. Fig.~\ref{Fig_data} shows one example of CXR image and corresponding ground truth. In terms of the example, semantic segmentation on CXR images is to classify pixels in the original image into two classes: Lung (white region in ground truth) and NonLung (black region in ground truth).

We split the dataset into training (70\% data), validation (10\% data), and testing (20\% data) datasets. 
%, where the detailed sample distributions in these three datasets are shown in Table~\ref{tab_data}.
 
%\begin{table}[h!]
%	\caption{Sample distribution in different classes for training, validation, and testing datasets}
%        \begin{center}
%                \begin{tabular}{|l|ccc|}
%                \hline \textbf{Dataset}  & \textbf{COVID-19} & \textbf{Non-COVID-19} & \textbf{Total} \\ \hline
%                           Training		& 362	& 4,120  	&   4,482	 \\    
%                           Validation	& 52		& 588  	&   640	 \\    
%                           Testing		& 103	& 1,177	&   1,280	\\
%                           Total		& 517	& 5,885	&   6,420	\\
%                   \hline        
%                \end{tabular}
%       \end{center}
%       \label{tab_data} 
%\end{table}

\subsection{Experimental settings}

%We employ fully convolutional network (FCN)~\cite{long2015fully} as baseline for COVID-19 image segmentation. In addition, it is enhanced through integrating ResNet~\cite{he2016deep} and MobileNets~\cite{howard2017mobilenets}. Moreover, two state-of-the-art methods are introduced for evaluation, namely, UNet~\cite{ronneberger2015u} and PSPNet~\cite{zhao2017pyramid}. 

We employed five state-of-the-art individual models as baselines to evaluate performance of semantic segmentation, namely, UNet~\cite{ronneberger2015u}, PSPNet~\cite{zhao2017pyramid}, FCN32~\cite{long2015fully} (FCN with 32$\times$upsampling), FCN32\_ResNet50 (FCN32 combined with ResNet50~\cite{he2016deep}), FCN32\_MobileNet(FCN32 combined with MobileNet~\cite{howard2017mobilenets}), and an ensemble baseline built based on majority voting, where the ensemble baseline is built based on bagging learning with these results generated by these five baselines (UNet, PSPNet, FCN32, FCN32\_ResNet50, and FCN32\_MobileNet). Moreover, key hyper-parameters of these individual models are shown in Table~\ref{tab_hp}.

\begin{table}[ht!]
	\caption{Hyper-parameters of baselines for COVID-19 image segmentation. }
\begin{center}
                \begin{tabular}{|l|ccc|}
                \hline \textbf{Model} 				& \textbf{Learning Rate} & \textbf{Batch Size} & \textbf{Epoch}   \\ \hline 
                UNet 							& 1e-3 & 2   & 50   \\ \hline
                PSPNet						& 1e-3 & 2   & 70  \\ \hline
                FCN32						& 1e-3 & 2   & 50    \\ \hline
                FCN32\_ResNet50 (F32\_R50)  	& 1e-3 & 2   & 50    \\ \hline
                FCN32\_MobileNet (F32\_M)  		& 1e-3 & 2   & 50    \\ \hline
 		  	                      
                \end{tabular}
       \end{center}
         \label{tab_hp}
\end{table}

We implemented two versions of the proposed approach including Ensemble (Weighted Voting (ECE)) and Ensemble (Weighted Voting (MCE)).  Ensemble (Weighted Voting (ECE)) is a weighted bagging learning method, where the weights are obtained by calculating expected calibration error (ECE). Similarly,  Ensemble (Weighted Voting (MCE)) is a weighted bagging learning method, where the weights are obtained by calculating maximum calibration error (MCE). Moreover, we combine the predictions of Ensemble (Majority Voting (MV)), Ensemble (Weighted Voting (ECE)), and Ensemble (Weighted Voting (MCE)) by majority voting to build Ensemble (Majority Voting + ECE + MCE (MVEM)). 

%Finally, we built an ensemble model based bagging learning as a baseline.
\begin{table*}[ht!]
	\caption{ Comparing performance between the baselines and the proposed method based on various evaluation metrics and corresponding standard deviations. }
       
        \begin{center}
                \begin{tabular}{|l|cccccc|}
                \hline \textbf{DL} 				& \textbf{Accuracy (\%)} & \textbf{Sensitivity (\%)} & \textbf{Specificity  (\%)} & \textbf{F1score  (\%)}& \textbf{ECE  (\%)} & \textbf{MCE  (\%)}  \\ \hline 
                		  	UNet											&  95.4\textpm2.5		& 90.7\textpm3.9		& 88.9\textpm4.5 		& 93.4\textpm3.0		&  3.2\textpm1.3 	&  39.7\textpm18.8	\\ 
                	           	PSPNet										&  95.0\textpm2.0		& 89.1\textpm3.9 		& 88.2\textpm4.3 		& 92.5\textpm2.9 		&  4.6\textpm1.2  	&  40.6\textpm14.9 	\\
				FCN32 										&  95.8\textpm2.4		& 92.3\textpm4.5		& 91.0\textpm5.0 		& 94.0\textpm3.5		&  2.5\textpm2.1 	&  37.6\textpm19.1 	\\ 
                	           	FCN32\_ResNet50 (F32\_R50) 					&  96.0\textpm2.5		& 92.3\textpm5.5 		& 91.4\textpm5.9 		& 94.3\textpm3.8 		&  2.3\textpm2.3  	&  29.8\textpm20.3 	\\
                 	   	FCN32\_MobileNet (F32\_M)						&  95.2\textpm2.3		& 91.0\textpm4.7		& 90.1\textpm5.6		& 93.1\textpm3.3 		&  4.1\textpm1.6  	&  38.2\textpm19.9 	\\	
			   	Ensemble (Majority Voting (MV))					&  98.8\textpm0.6 		& 94.1\textpm3.0 		& 92.9\textpm3.6 		& 96.6\textpm1.7 		&  2.4\textpm1.2 	&  28.1\textpm14.1 	\\
				\hline 
                 	   	Ensemble (Weighted Voting (ECE))					&  99.1\textpm0.5		& 95.4\textpm2.9  		& 94.3\textpm2.9		& 97.1\textpm1.5 		&  2.3\textpm1.2 	&  24.7\textpm12.4 	\\
			   	Ensemble (Weighted Voting (MCE))					&  98.7\textpm0.7 		& 93.9\textpm3.7 		& 92.6\textpm3.7  		& 96.3\textpm1.9		&  2.4\textpm1.2   	&  28.9\textpm14.6 	\\
				Ensemble (Majority Voting + ECE + MCE (MVEM))		&  99.2\textpm0.4		& 97.7\textpm2.3 		& 95.4\textpm2.3 		& 98.4\textpm0.8  		&  2.1\textpm1.1  	&  20.1\textpm10.1	\\
				 \hline
                      
                \end{tabular}
       \end{center}
         \label{tab_cp}
\end{table*}

\begin{figure*}[h!]
 \begin{center}
\begin{tabular}{ccccc}
	\photo{Figure/result/r1/original.png} &  \photo{Figure/result/r1/gt.png} &  \photo{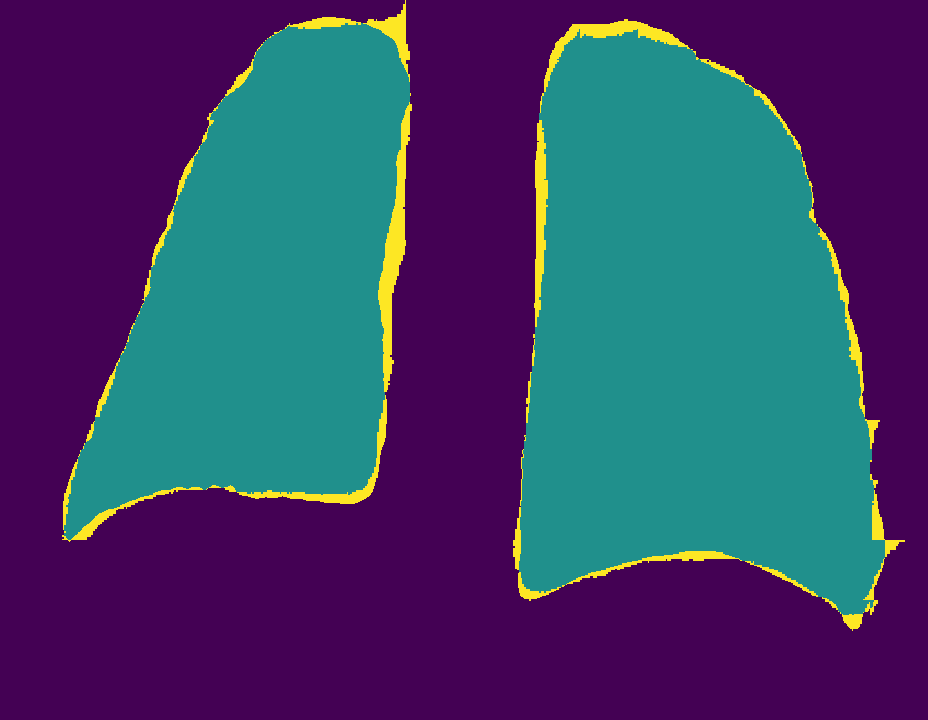} & \photo{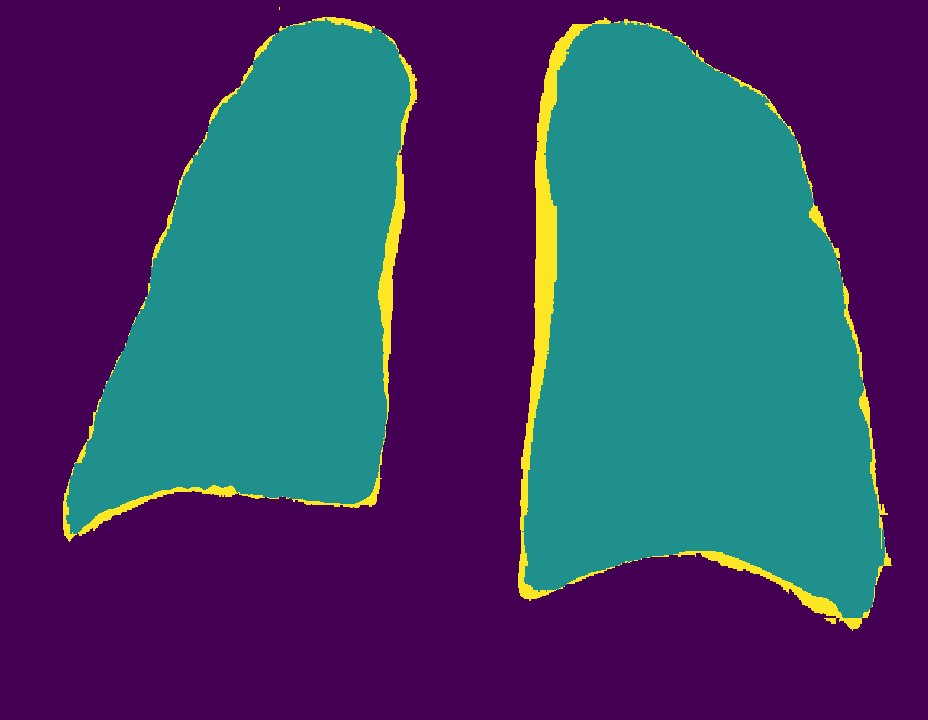} &  \photo{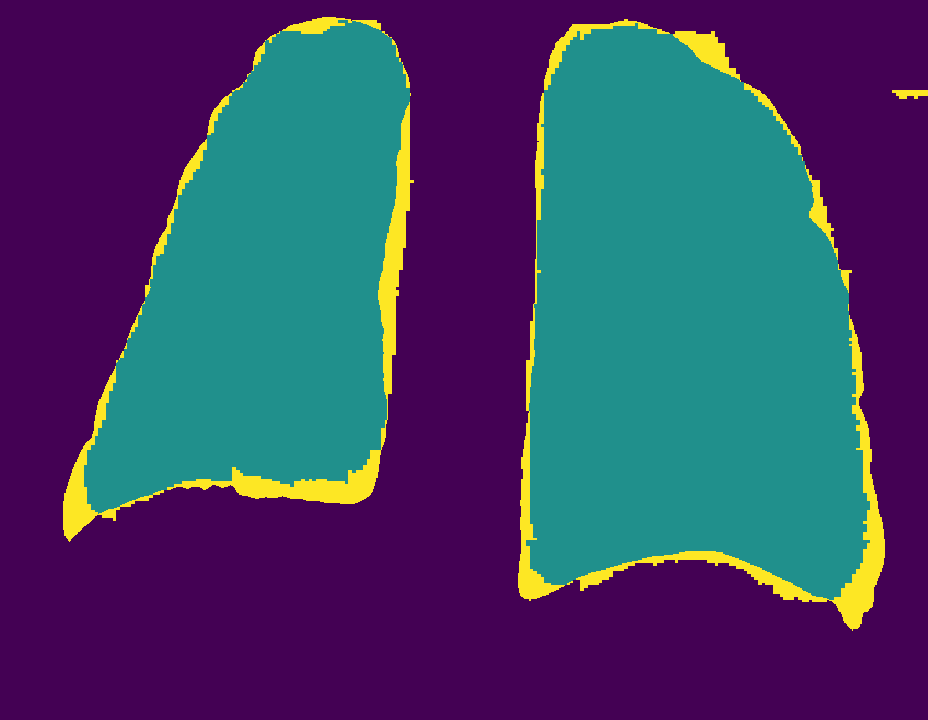}\\
	(a) original image &  (b) ground truth  &  (c) FCN & (d) F32\_R50 &  (e) F32\_M   \\
	  \photo{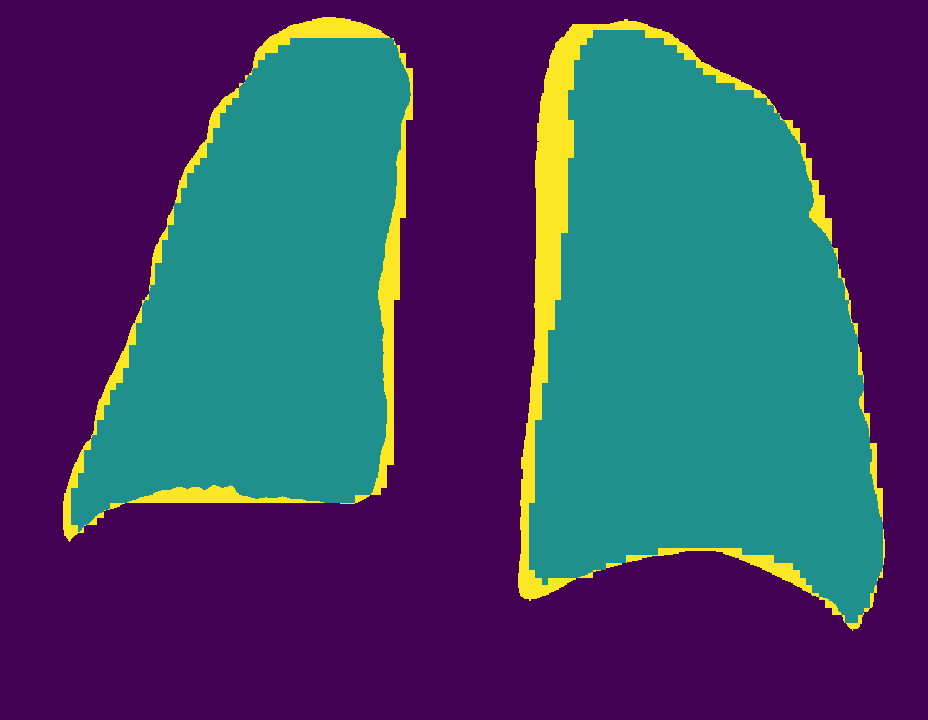} &\photo{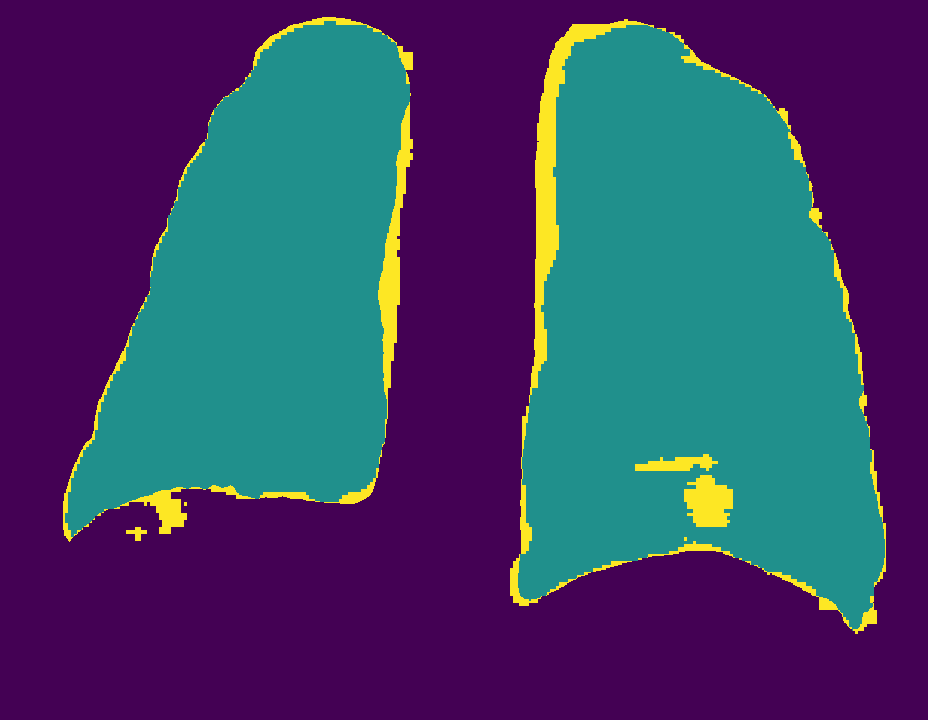} &  \photo{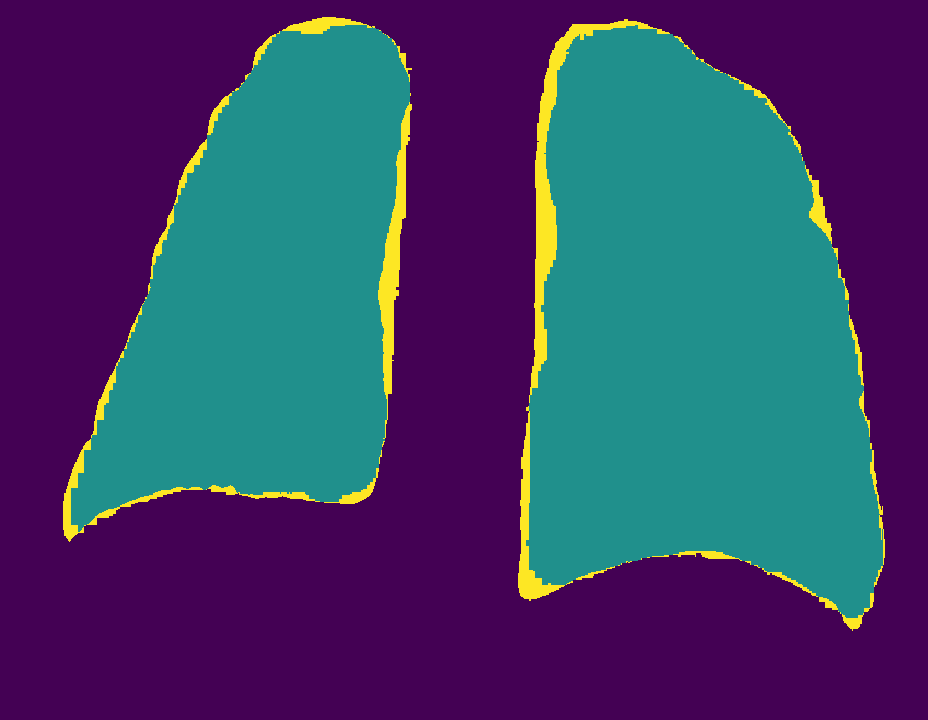} &  \photo{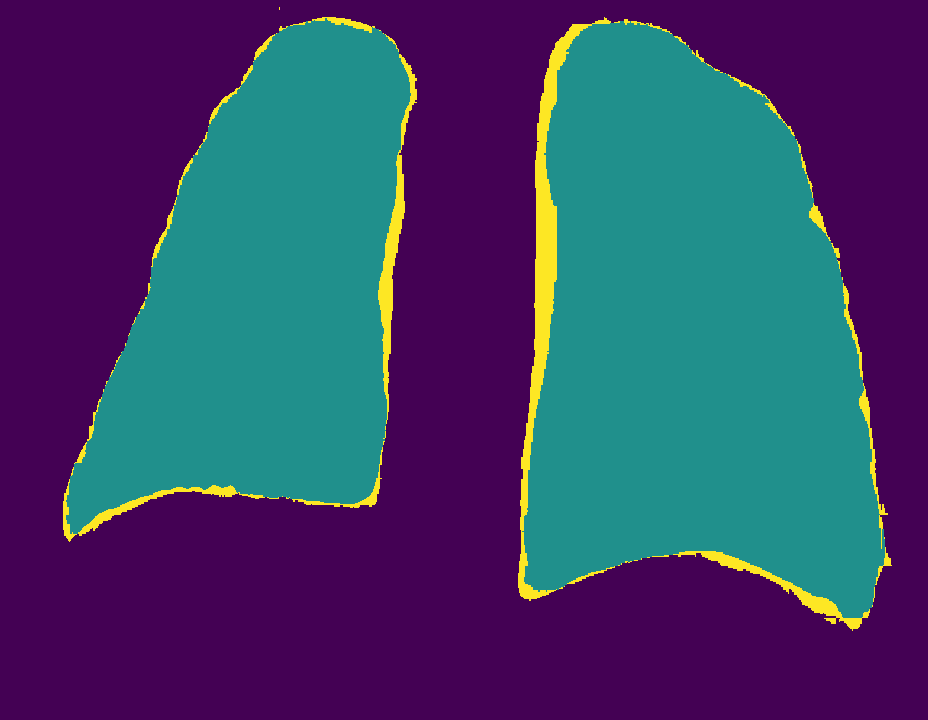} &  \photo{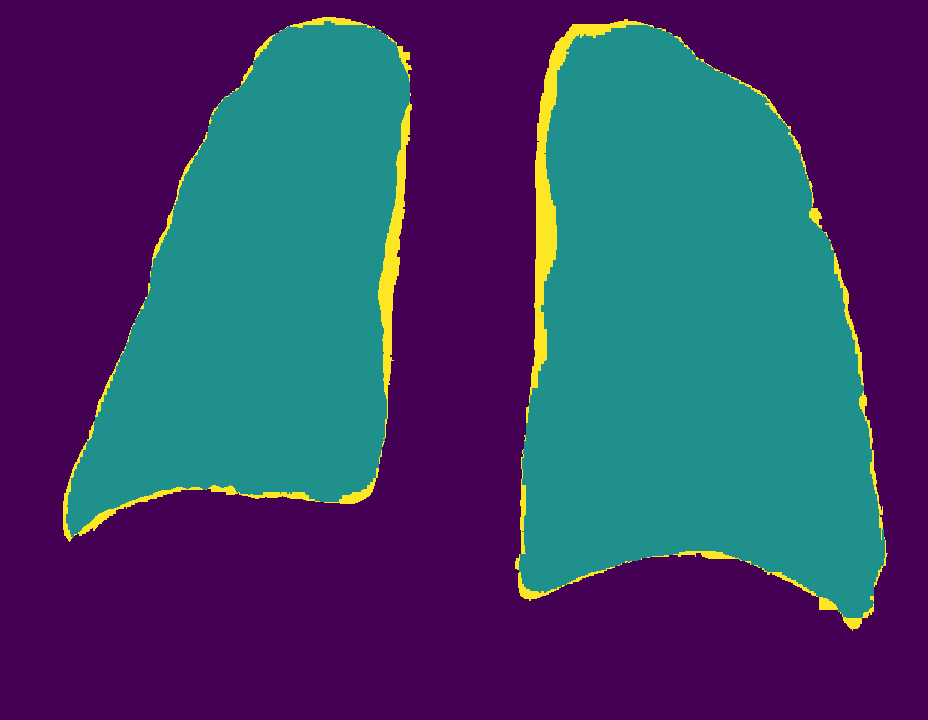}\\
	(f) PSNet & (g) UNet &  (h) MV   &  (i) ECE &  (j) MVEM  \\	
\end{tabular}
 \end{center}
 \caption{An example of prediction visualization on semantic segmentation generated by the baselines and proposed models. F32\_R50 and  F32\_M denotes FCN32\_ResNet50 and FCN32\_MobileNet while MV, ECE, and MVEM denotes  Ensemble (Majority Voting (MV)), Ensemble (Weighted Voting (ECE)), and Ensemble (Majority Voting + ECE + MCE (MVEM)). In the predictions, purple color, yellow color, and green color denotes background, incorrect prediction, and correct prediction, respectively, where the smaller region of yellow color means higher performance.}
 \label{Fig_covid-19}
\end{figure*}

\subsection{Evaluation metric}
Various evaluation metrics are employed to evaluate the performance of our proposed model, which includes accuracy, F1score, sensitivity, and specificity. Accuracy is calculated by dividing the number of pixels identified correctly over the total number of pixels in chest X-ray images. 

\begin{equation}
	Accuracy = \frac{N_{correct}}{N_{total}}.
\end{equation}

\begin{equation}
	Fscore = \frac{2 \times Precision \times Recall}{Precision + Recall}.
\end{equation}

where $Precision$ defines the capability of a model to represent only correct pixels and $Recall$ computes the aptness to refer all corresponding correct pixels.

\begin{equation}
	Precision = \frac{TP}{TP+FP}.
\end{equation}

\begin{equation}
	Recall = \frac{TP}{TP+FN}.
\end{equation}
whereas ${TP}$ (True Positive) counts the total number of pixels that matches the annotated pixels of RIOs. ${FP}$ (False Positive) measures the number of pixels that don't belong to RIOs, but are recognized as pixels of RIOs. ${FN}$ (False Negative) counts the number of pixels of RIOs are recognized as those don't belong to RIOs. 
The main goal for binary classification is to improve the recall without hurting the precision. However, recall and precision goals are often conflicting, since when increasing the true positive (TP) for the minority class (True), the number of false positives (FP) can also be increased; this will reduce the precision~\cite{chawla2009data}.

Moreover, we employed  sensitivity and specificity to evaluate performance of semantic segmentation~\cite{saood2021covid}, where the sensitivity measures how good a test is at detecting the RIOs while the specificity refers to  how good a test is at avoiding false alarms.

\begin{equation}
	Specificity = \frac{TN}{TN+FP}.
\end{equation}

\begin{equation}
	Sensitivity = \frac{TP}{TP+FN}.
\end{equation}

whereas ${TN}$ (True Negative) counts total number of pixels that don't belong to RIOs and are recognized as those don't belong to RIOs.

Finally, we employ expected calibration error (ECE)\footnote{\url{https://www.tensorflow.org/probability/api_docs/python/tfp/stats/expected_calibration_error}} and MCE to measurethe calibration errors~\cite{guo2017calibration} for evaluating the prediction uncertainty, where ECE and MCE are defined as equations (1) and (2), respectively. The lower ECE and MCE are, the higher prediction certainty is.

\subsection{Experimental results}

We validate the proposed method from two perspectives: comprehensive performance comparison between the baselines and the proposed method, and hyper-parameter examination.

\begin{table*}[ht!]
	\caption{ Comparing performance of the proposed methods built with different number of individual models. }
       
        \begin{center}
                \begin{tabular}{|l|cccccc|}
                
                \hline \multicolumn{7}{|c|}{Ensemble 2 (FCN32\_RESNET50 + FCN32) }\\  \hline
                \textbf{DL} & \textbf{Accuracy (\%)} & \textbf{Sensitivity (\%)} & \textbf{Specificity  (\%)}  & \textbf{F1score  (\%)} & \textbf{ECE  (\%)} & \textbf{MCE  (\%)}  \\ \hline 
				
                	           	FCN32\_ResNet50 (F32\_R50) 					&  95.8\textpm2.1		& 92.3\textpm3.9		& 91.0\textpm4.5 		& 94.0\textpm3.0		&  2.5\textpm1.3 	&  37.6\textpm18.8 	\\

                 	   	Ensemble (Weighted Voting (ECE))					&  99.0\textpm0.5 		& 95.3\textpm2.4  		& 94.4\textpm2.8		& 96.9\textpm1.6 		&  2.3\textpm1.2  	&  22.3\textpm11.3 	\\
			   	Ensemble (Weighted Voting (MCE))					&  98.8\textpm0.6 		& 93.8\textpm3.1 		& 93.7\textpm3.2  		& 96.4\textpm1.8		&  2.5\textpm1.3   	&  25.1\textpm12.6 	\\
				\hline 
				\hline \multicolumn{7}{|c|}{Ensemble 3 (FCN32\_RESNET50 + FCN32 + UNET) }\\  \hline
                \textbf{DL} & \textbf{Accuracy (\%)} & \textbf{Sensitivity (\%)} & \textbf{Specificity  (\%)}  & \textbf{F1score  (\%)} & \textbf{ECE  (\%)} & \textbf{MCE  (\%)}  \\ \hline  
				Ensemble (Majority Voting (MV))					&  98.7\textpm0.7 		& 93.9\textpm3.0 		& 94.1\textpm3.0 		& 96.1\textpm2.0 		&  2.7\textpm1.4 	&  31.1\textpm15.6 	\\
                 	   	Ensemble (Weighted Voting (ECE))					&  98.4\textpm0.8 		& 95.5\textpm2.3  		& 94.9\textpm2.6		& 96.9\textpm1.6 		&  2.8\textpm1.4	&  26.1\textpm13.1 	\\
			   	Ensemble (Weighted Voting (MCE))					&  98.3\textpm0.9 		& 93.1\textpm3.5 		& 93.0\textpm3.5  		& 96.0\textpm2.0 		&  2.8\textpm1.4   	&  31.2\textpm15.6 	\\
				Ensemble (Majority Voting + ECE + MCE (MVEM))		&  98.8\textpm0.6 		& 97.6\textpm1.2 		& 96.4\textpm1.8  		& 98.1\textpm1.0  		&  2.1\textpm1.1   	&  21.1\textpm10.6 	\\
				 \hline
				 \hline \multicolumn{7}{|c|}{Ensemble 4 (FCN32\_RESNET50 + FCN32 + UNET + FN32\_MOBILENET) }\\  \hline
               \textbf{DL} & \textbf{Accuracy (\%)} &\textbf{Sensitivity (\%)} & \textbf{Specificity  (\%)} & \textbf{F1score  (\%)} & \textbf{ECE  (\%)} & \textbf{MCE  (\%)}  \\ \hline  
				Ensemble (Weighted Voting (ECE))					&  98.3\textpm0.9 		& 95.0\textpm2.5  		& 94.6\textpm2.7		& 96.5\textpm1.8		&  2.7\textpm1.4  	&  24.3\textpm13.5 	\\
			   	Ensemble (Weighted Voting (MCE))					&  97.9\textpm1.1		& 94.1\textpm3.0 		& 93.8\textpm3.1  		& 96.1\textpm2.0 		&  3.0\textpm1.5   	&  32.3\textpm15.0	\\

				 \hline
				 \hline \multicolumn{7}{|c|}{Ensemble 5 (FCN32\_RESNET50 + FCN32 + UNET + FCN32\_MOBILENET + PSPNET) }\\  \hline
                \textbf{DL} & \textbf{Accuracy (\%)} & \textbf{Sensitivity (\%)} & \textbf{Specificity  (\%)} & \textbf{F1score  (\%)} & \textbf{ECE  (\%)} & \textbf{MCE  (\%)}  \\ \hline  
				Ensemble (Majority Voting (MV))					&  98.8\textpm0.6 		& 94.1\textpm3.0 		& 92.9\textpm3.6 		& 96.6\textpm1.7 		&  2.4\textpm1.2 	&  28.1\textpm14.1 	\\       				    
				Ensemble (Weighted Voting (ECE))					&  99.1\textpm0.5		& 95.4\textpm2.9  		& 94.3\textpm2.9		& 97.1\textpm1.5 		&  2.3\textpm1.2 	&  24.7\textpm12.4 	\\
			   	Ensemble (Weighted Voting (MCE))					&  98.7\textpm0.7 		& 93.9\textpm3.7 		& 92.6\textpm3.7  		& 96.3\textpm1.9		&  2.4\textpm1.2   	&  28.9\textpm14.6 	\\
				Ensemble (Majority Voting + ECE + MCE (MVEM))		&  99.2\textpm0.4		& 97.7\textpm2.3 		& 95.4\textpm2.3 		& 98.4\textpm0.8  		&  2.1\textpm1.1  	&  20.1\textpm10.1	\\ \hline
                \end{tabular}
       \end{center}
         \label{tab_hyper}
\end{table*}

\begin{figure}[h!]
 \begin{center}
\begin{tabular}{ccc}

	\photo{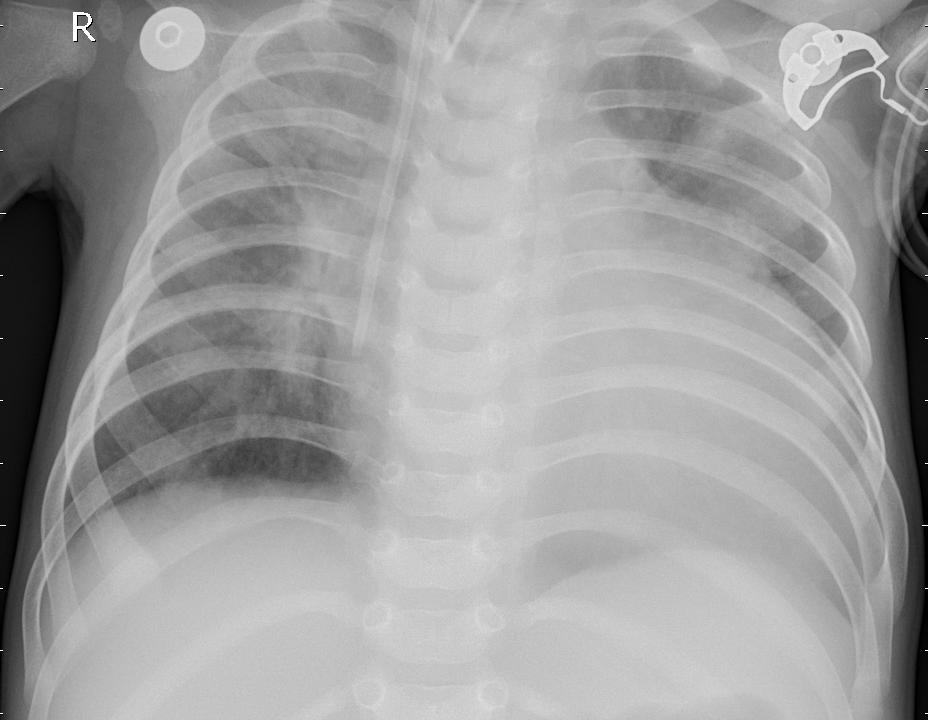} &  \photo{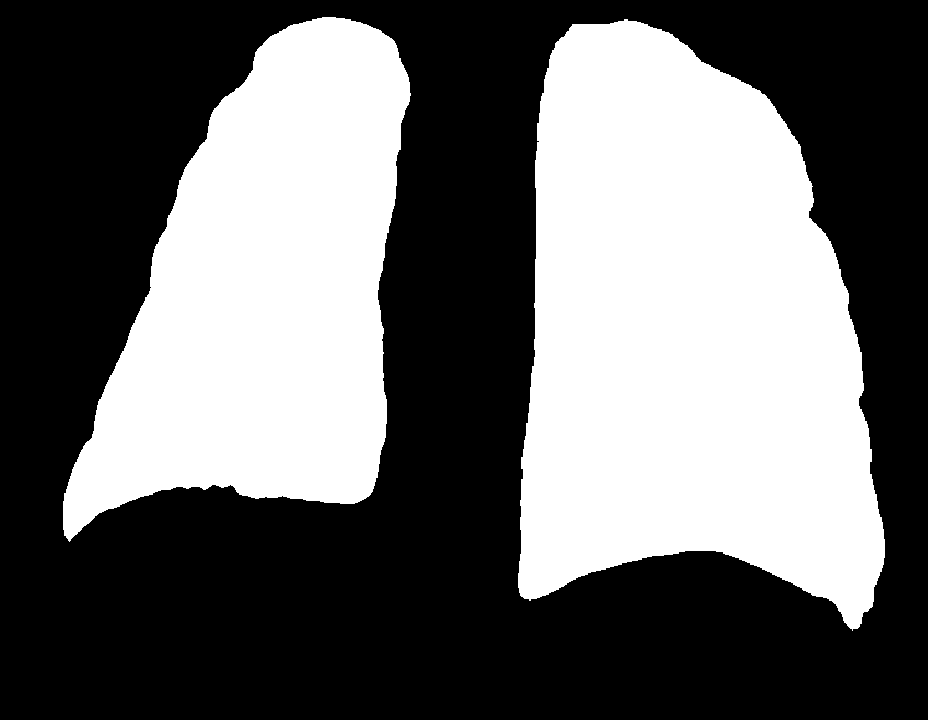} &  \photo{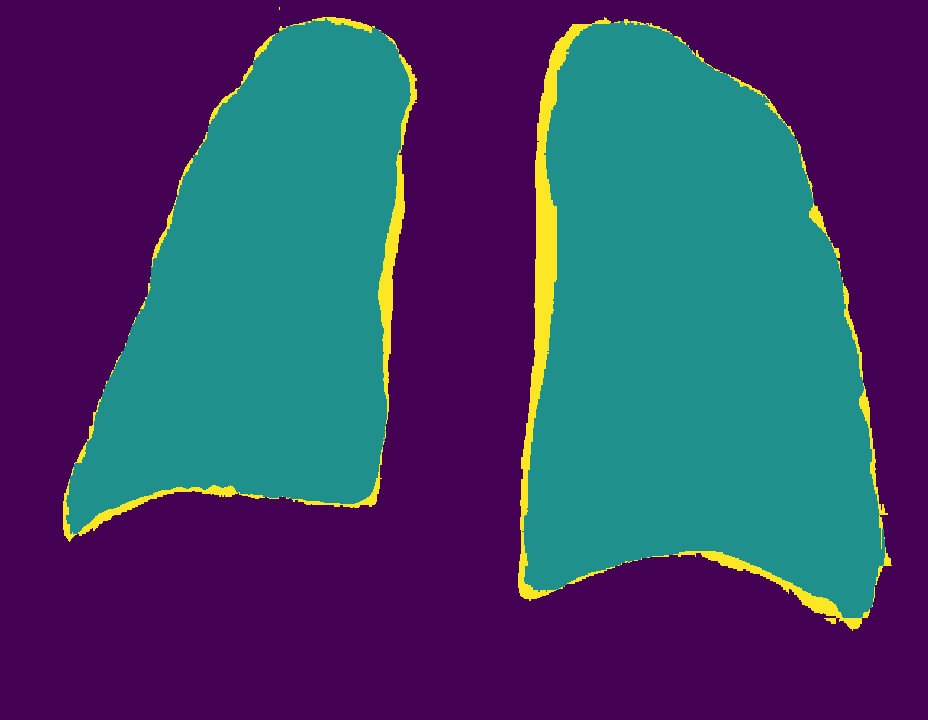} \\
	(a) original image &  (b) ground truth  &  (c) F32\_R50    \\ 
	%\multicolumn{3}{|c|}{Ensemble 3}\\  \hline
	  \photo{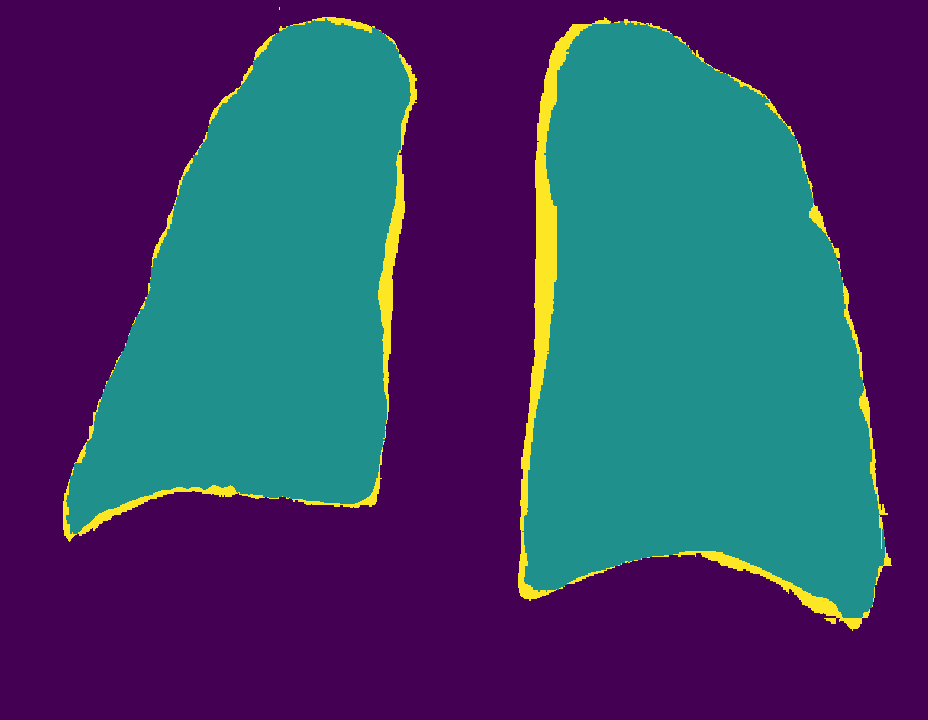} &\photo{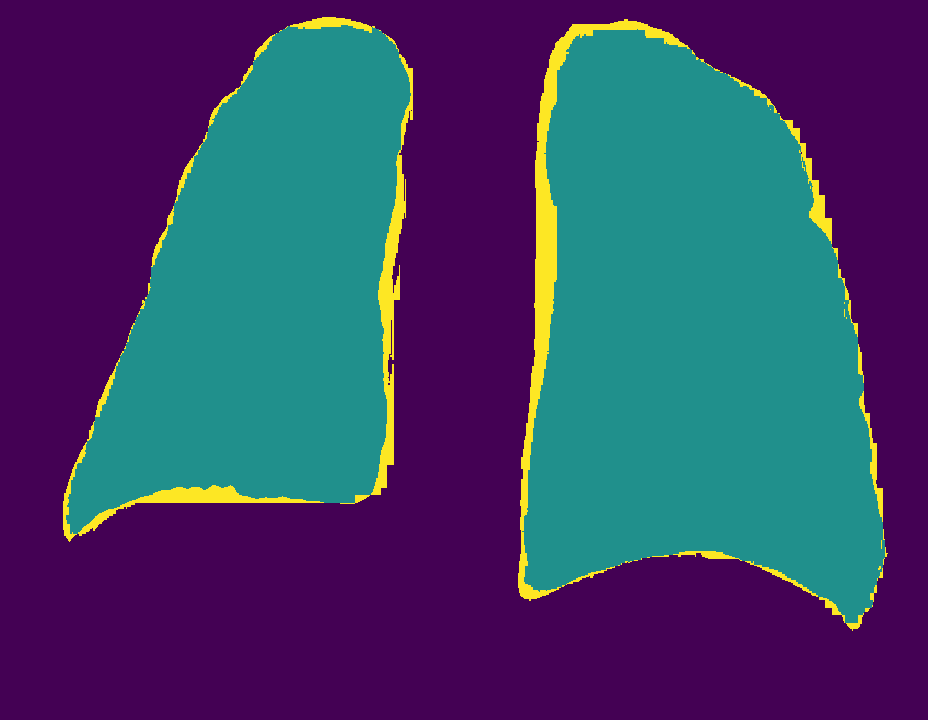}  &  \photo{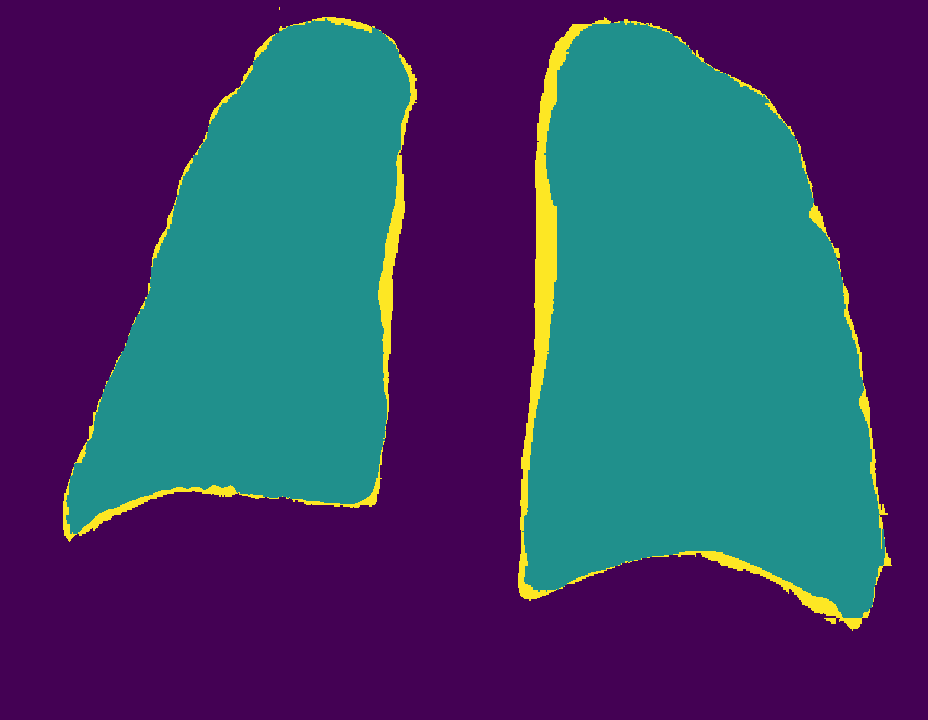}\\
	(d) ECE\_3 & (e) MCE\_3 &  (f) MVEM\_3     \\	
	%\multicolumn{3}{|c|}{Ensemble 5}\\  \hline
	 \photo{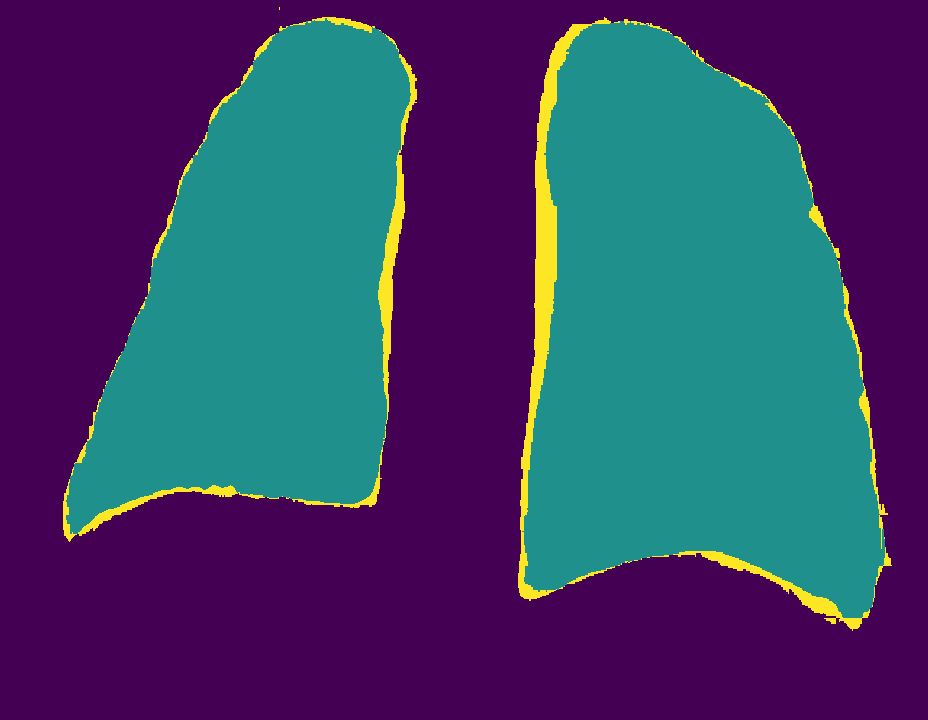} &\photo{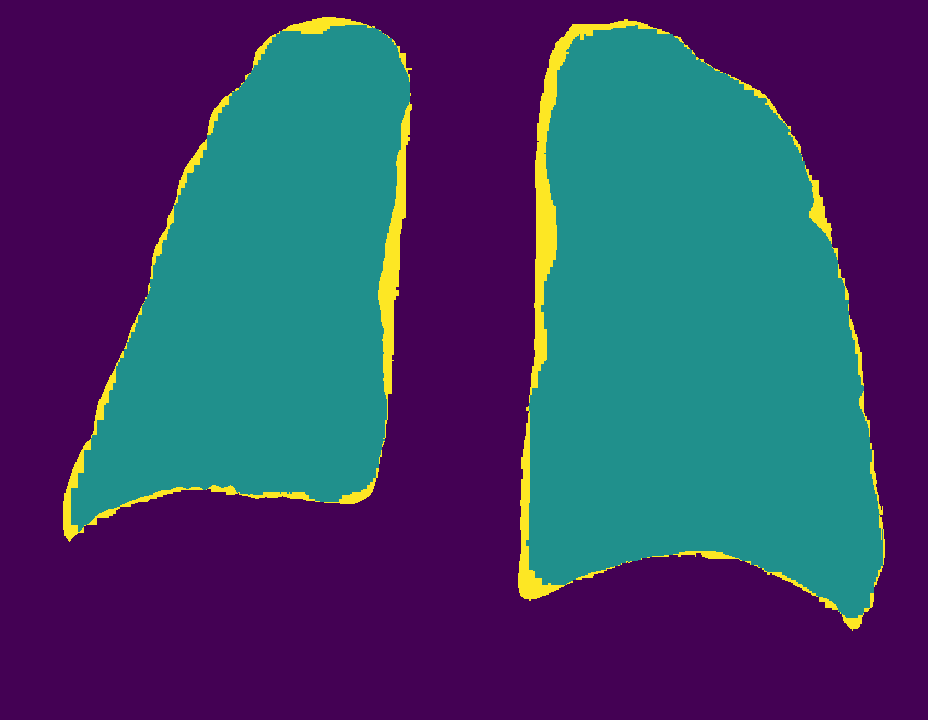}  &  \photo{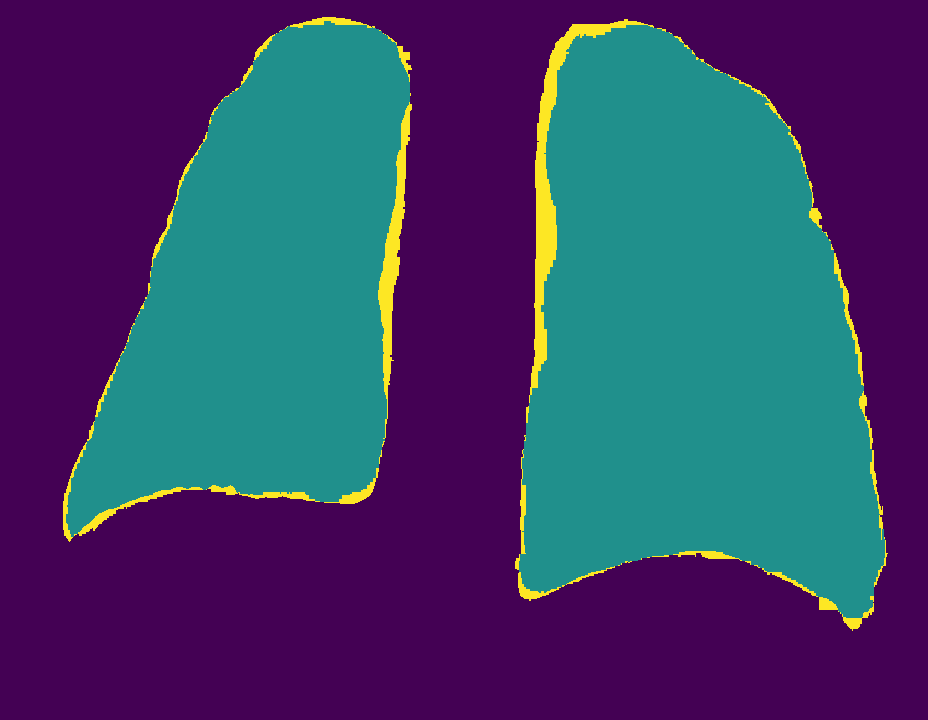}\\
	(g) ECE\_5 & (h) MCE\_5 &  (i) MVEM\_5       \\	

\end{tabular}
 \end{center}
 \caption{Comparison of prediction visualization produced by the proposed methods built with different number of individual models. The second row presents the predictions generated by three implementations of the proposed method with Ensemble 3 (FCN32\_RESNET50 + FCN32 + UNET) including ECE\_3 (Ensemble (Weighted Voting (ECE))), MCE\_3 (Ensemble (Weighted Voting (MCE))), and MVEM\_3 (Ensemble (Majority Voting + ECE + MCE (MVEM))). Similarly, The thrid row presents the predictions generated by five implementations of the proposed method with Ensemble 5 (FCN32\_RESNET50 + FCN32 + UNET + FCN32\_MOBILENET + PSPNET)  including ECE\_5, MCE\_5, and MVEM\_5. In the predictions, purple color, yellow color, and green color denotes background, incorrect prediction, and correct prediction, respectively, where the smaller region of yellow color means higher performance.}
 \label{Fig_hyper}
\end{figure}

\subsubsection{Performance Comparison}
Table~\ref{tab_cp} presents the performance comparison between the state-of-the-art individual models  and the proposed method in terms of various evaluation metrics and corresponding standard deviations. We can observe that these individual models can perform well on COVID-19 image segmentation regarding F1scores and Accuracy. Moreover, prediction uncertainties of most of them are promising with respect to ECE and MCE.  For these individual models, FCN32\_ResNet50 outperforms other individual models with higher certainty. In addition, as one baseline, Ensemble (Majority Voting (MV)) performs better than other individual methods with highest prediction certainty by comparing F1score, ECE and MCE. It means that combining predictions of these individual models can effectively improve performance and prediction certainty.

For the proposed method, Ensemble (Weighted Voting (ECE)) can perform better than the baselines including these individual models and  Ensemble (Majority Voting (MV)) by comparing accuracy, recall, and F1score. Moreover, Ensemble (Weighted Voting (ECE)) is able to improve the prediction certainty. It means that using appropriate calibration errors as weights to implement weighted bagging deep learning can effectively improve prediction certainty as well as performance. In other words, it is an effective method to calibrate models by using appropriate calibration errors as weights to combine predictions. Furthermore, Ensemble (Majority Voting + ECE + MCE (MVEM)) obtains the optimal performance with highest prediction certainty. It indicates that ensemble strategy such as majority voting is effective to combine predictions to further improve performance and prediction certainty.  Moreover, Ensemble (Majority Voting + ECE + MCE (MVEM))  performed more stable since the standard deviations of performance and calibration errors are lower than those of baselines. %In terms of these observations, both the performance and prediction certainty for COVID-19 image segmentation can be enhanced through using calibration errors as weights to build weighted bagging learning.

In addition to the performance comparison, we show an example of prediction visualization on semantic segmentation generated by the baselines and proposed models in Fig.~\ref{Fig_covid-19}. When we examine the prediction visualization for these individual models, we can observe that they miss some key components (yellow regions) for detecting lung. Taking UNet as an example, through comparing the predictions with ground truth, key components highlighted with yellow color are missed on subfigure (g). On the contrary, ensemble models such as MV, ECE, and MVEM perform better in that regard of predictions since yellow regions in their predictions are smaller, where the proposed method including ECE and MVEM outperform other baselines. It means that the proposed method can effectively improve recall on detecting lung by distributing contributions of prediction based on calibration errors such as ECE and MCE.

\subsubsection{Hyper-parameter Examination}

Fine-tuning hyper-parameter for building deep learning models is an imperative step to obtain optimal performance. The process of building the proposed method involved various hyper-parameters. For example, for each individual DL model, we have to fine-tune learning rate, batch size, and epoch to achieve optimal performance. Specially, for the proposed bagging deep learning, how many individual models involved is still an open challenge. Here, we examine if the number of  individual models will significantly affect the performance of the proposed method.

Table~\ref{tab_hyper} presents the performance comparison for various bagging deep learning models built with different number of individual models. Generally speaking, more individual models will enhance performance and improve prediction certainty regarding F1score and ECE. When we employ five individual models (Ensemble 5 (FCN32\_RESNET50 + FCN32 + UNET + FCN32\_MOBILENET + PSPNET) ), we obtain the optimal performance and the highest prediction certainty regarding values of accuracy, F1score, and ECE for Ensemble (Weighted Voting (ECE)) and Ensemble (Majority Voting + ECE + MCE (MVEM)), where the values of Recall and F1score are improved more significantly than other evaluation metrics. 

Additionally, Fig.~\ref{Fig_hyper} shows comparison of prediction visualization produced by the proposed methods built with different number of individual models. It is observed that more individual models involved in the proposed approach will reduce the size of missing components. Moreover, MVEM outperforms other ensemble methods, which means that majority voting based on more individual DL models can further enhance the performance of recognition of RIOs.

In summary, in terms of observations mentioned above, the proposed method can effectively improve semantic segmentation, as well as reduce the prediction uncertainty through using the calibration error as weights of DL models to combine their predictions. Moreover, more individual DL models involved in the implementation of the proposed approach can further enhance the performance and prediction certainty, which meets the intuition of majority voting for bagging deep learning. To some extent, it is an effective method to combine advantages of these individual DL models to improve the task performance without complex implementations.

%% file: Relatedwork.tex
This paper aims to build a novel bagging learning method to implement COVID-19 semantic segmentation through combining bagging deep learning and model calibration. Semantic segmentation has achieved significant successes by developing deep learning models such as U-Net~\cite{ronneberger2015u} and V-Net~\cite{milletari2016v}. In the biomedical domain, there have been numerous techniques for lung segmentation with different purposes~\cite{cciccek20163d, isensee2018nnu}. The U-Net is an effective technique for segmenting both lung regions and lung lesions in COVID applications~\cite{cao2020longitudinal}. The U-Net built with fully convolutional network~\cite{ronneberger2015u} has a U-shape architecture with two symmetric paths: encoding path and decoding path. The layers at the same level in two paths are connected by the shortcut connections, which is to learn better visual semantics as well as detailed contexture. Zhou \textit{et al.}~\cite{zhou2018unet++} proposed the UNet++ that inserts a nested convolutional structure between the encoding and decoding path. In addition, Milletari \textit{et al.}~\cite{milletari2016v} built V-Net using the residual blocks as the basic convolutional block, and optimized the network by a Dice loss. Furthermore, Shan \textit{et al.}~\cite{shan2020lung} built VB-Net for more efficient segmentation by equipping the convolutional blocks with the so-called bottleneck blocks.  Moreover, U-Net and its variants have been developed, achieving reasonable segmentation results in COVID-19 diagnosis~\cite{chen2020deep}. In recent years, attention mechanisms can learn the most discriminant part of the features in deep learning models. Oktay \textit{et al.}~\cite{oktay2018attention} proposed an Attention U-Net  to capture fine structures in medical images, thereby suitable for segmenting lesions and lung nodules in COVID-19 applications.  

%Deep learning has promoted the development in various domains such as computer vision (CV)~\cite{krizhevsky2012imagenet, he2016deep} and natural language process (NLP)~\cite{devlin2018bert}. Furthermore, it has been applied to more and more 
Safety-critical applications like medical image processing~\cite{esteva2017dermatologist}, autonomous driving~\cite{bojarski2016end},  and precipitation forecasting~\cite{sonderby2020metnet} not only require high accuracy, but also need high prediction uncertainty measured by the model calibration. Two categories of methods are proposed to calculate the model calibration, namely, Bayesian-based and Non-Bayesian-based. Bayesian-based methods refer to Bayesian neural networks that estimates prediction/model uncertainty based on Bayesian process. The main concern of such methods is associate with its high computation complex and prior assumption on model weights. To reduce the computation complexity and enhance the scalability of Bayesian neural networks for data analysis on larger datasets, Hern{\'a}ndez-Lobato \textit{et al.}~\cite{hernandez2015probabilistic} proposed probabilistic back-propagation for learning Bayesian neural networks. Non-Bayesian-based methods develop various strategies such as model ensemble~\cite{lakshminarayanan2016simple} and prior assumption on predictions~\cite{sensoy2018evidential} to estimate the prediction uncertainty, which is to reduce the cost of estimating the uncertainty. To reduce computation cost and training difficulty, Lakshminarayanan \textit{et al.}~\cite{lakshminarayanan2016simple} proposed deep ensemble that is simply to implement, trained in a parallel manner, requires less hyper-parameter tuning, and estimates high quality predictive uncertainty. However, it is very tricky to obtain the optimal number of individual models to build deep ensemble for various applications. Moreover, to reduce the cost of the memory usage and inference of Bayesian neural networks and deep ensembles, Liu (2020) \textit{et al.}~\cite{liu2020simple} proposed approaches to estimate uncertainty by building only one neural networks with two steps: 1) Measuring the distance between testing samples and  training samples; 2) Implementing spectral-normalized neural Gaussian process (SNGP) that is to improve the measurement of the distance by adding a weight normalization step during training and replacing the output layer with a Gaussian process. However, experimental results on dialog intent detection indicated that deep ensemble performed better than the proposed method on many evaluation metrics such as accuracy. Recently, Wilson \textit{et al.}~\cite{wilson2020case} systematically summarized Bayesian deep learning and claimed that deep ensemble can be treated as approximate Bayesian marginalization of model parameters. On the other side, they also claimed that Bayesian methods were not perfect regarding prior assumptions on model weights. %To address the concern on prior assumption on distribution of model weights for Bayesian deep learning, Sensoy \textit{et al.}~\cite{sensoy2018evidential} proposed to explicitly modeling the class probability based on Dirichlet distribution rather than estimating the class uncertainty in terms of prior assumptions of model weights. This proposed method was validated on image classification with MNIST and CIFAR5 datasets. The results, however, seems not to demonstrate significant improvement on performance by comparing Dropout and Deep Ensemble~\cite{lakshminarayanan2016simple}. 

In terms of previous work on model calibration and semantic segmentation, we proposed the calibrated ensemble model to not only enhance performance on semantic segmentation, but also reduce the prediction uncertainty.

%% file: Conclusion.tex
In this paper, a novel bagging deep learning model is proposed for COVID-19 image segmentation on chest x-ray images. It combines the model calibration and traditional bagging learning to not only enhance the segmentation performance, but also improve the prediction certainty that is extremely important to high-risk applications in biomedical domain. We validate the proposed method on a large chest x-ray dataset that is associated with COVID-19. Experimental results demonstrate that the proposed model could recognize the lung region more effectively through comparing with state-of-the-art baselines. For the future work, we plan to extend the proposed model for building an end-to-end model for both COVID-19 image classification and image segmentation.